\documentclass[twocolumn,,nofootinbib,preprintnumbers,superscriptaddress]{revtex4}

\usepackage[dvipdf,dvips,dvipdfmx]{graphicx}
\usepackage{amsmath}
\usepackage{amssymb}
\usepackage{float}
\usepackage{wrapfig}
\usepackage{bm}

\newcommand{\Slash}[1]{{\ooalign{\hfil/\hfil\crcr$#1$}}} 
\newcommand{\bvec}[1]{\mbox{\boldmath $#1$}}

\newcommand{\al}[1]{\begin{align}#1\end{align}}
\newcommand{\bp}{\begin{pmatrix}}
\newcommand{\ep}{\end{pmatrix}}

\newcommand{\del}{\partial}

\newcommand{\bs}[1]{\boldsymbol}

\begin{document}

\title{Functional renormalization group study of the Nambu--Jona-Lasinio model at finite temperature and density in an external magnetic field}

\author{Ken-Ichi \surname{Aoki}}
\email{aoki@hep.s.kanazawa-u.ac.jp}
\affiliation{Institute for Theoretical Physics, Kanazawa University, Kanazawa 920-1192, Japan}

\author{Hidenari \surname{Uoi} \footnote{H. U. belonged to the Institute until March 2014.}}
\email{huoi@hep.s.kanazawa-u.ac.jp}
\affiliation{Institute for Theoretical Physics, Kanazawa University, Kanazawa 920-1192, Japan}

\author{Masatoshi \surname{Yamada}}
\email{masay@hep.s.kanazawa-u.ac.jp}
\affiliation{Institute for Theoretical Physics, Kanazawa University, Kanazawa 920-1192, Japan}

\preprint{KANAZAWA-15-12}

\begin{abstract}
In this study, we investigate the Nambu--Jona-Lasinio (NJL) model at finite temperature and finite density in an external magnetic field using the functional renormalization group. 
We investigate the dependence of the position of the ultraviolet fixed point (UVFP) of the four-Fermi coupling constant on the temperature, density, and external magnetic field, and we obtain the chiral phase structure.
The UVFP at low temperature and finite chemical potential oscillates in a small external magnetic field, which can be interpreted as the de Haas--van Alphen effect.
We also obtain phase diagrams with complex structures, where the phase boundary moves back and forth as the external magnetic field increases in the low temperature and high density region.
\end{abstract}
\maketitle

\section{Introduction}
The phase diagram and equation of state for quantum chromodynamics (QCD) have been studied as important subjects in the area of elementary particle physics.
In particular, QCD matter in an external magnetic field has attracted much attention for several reasons.
It is known that neutron stars, which are high baryonic density stars, generate strong external magnetic fields ($B\sim 10^{14}$ Gauss)~\cite{Duncan:1992hi}.
In heavy-ion collision experiments, quite strong external magnetic fields ($B\sim 10^{19}$~Gauss) are predicted to exist~\cite{Kharzeev:2007jp,Fukushima:2011jc,Fukushima:2008xe,Fukushima:2012zz,Fukushima:2012vr}.
Furthermore, it has been suggested that extremely strong magnetic fields ($B\sim 10^{23}$ Gauss) are generated via the cosmological electroweak phase transition~\cite{Vachaspati:1991nm} and quark-hadron phase transition~\cite{Cheng:1994yr}.
Thus, QCD matter in strong magnetic fields with $|eB|\sim \Lambda_{\rm QCD}^2$ is common in our universe.

The analysis of chiral effective models has shown that under an external magnetic field, the chiral symmetry is always broken due to dimensional reduction.
This phenomenon is called magnetic catalysis~\cite{Klimenko:1990rh,Klimenko:1991he,Gusynin:1994re,Gusynin:1994va,Gusynin:1994xp,Gusynin:1995nb,Shushpanov:1997sf} and the chiral restoration temperature increases with the strength of the magnetic field.
By contrast, the chiral restoration density decreases with a fixed low temperature~\cite{Preis:2010cq,Preis:2012fh,Fukushima:2012xw}. 

A recent lattice simulation~\cite{Bali:2011qj} of QCD with an external magnetic field indicated that the critical restoration temperature decreases with a stronger magnetic field, which is called inverse magnetic catalysis or magnetic inhibition~\cite{Fukushima:2012kc,Kojo:2012js,Bruckmann:2013oba,Chao:2013qpa,Fraga:2013ova}. 
The analysis presented in the present study does not provide an explanation of this effect, but instead it might suggest that the physics responsible for the magnetic inhibition at high temperature is beyond the physics captured by the Nambu--Jona-Lasinio (NJL) model.

In this study, we analyze the NJL model, which describes dynamical chiral symmetry breaking (D$\chi$SB)~\cite{Nambu:1961tp,Nambu:1961fr,Hatsuda:1994pi}, in a strong magnetic field at finite temperature and finite density using the functional renormalization group (FRG)~\cite{Aoki:2000wm,Berges:2000ew,Kodama:1999if,Pawlowski:2005xe,Gies:2006wv,Braun:2011pp}.
The basic properties of this system such as the renormalization group (RG) flow equations and the fixed point structure have been investigated in many studies~\cite{Aoki:1999dv,Fukushima:2012xw,Scherer:2012nn,Andersen:2012bq,Andersen:2013swa,Costa:2013zca,Allen:2013lda,Ferreira:2013tba,Ferreira:2013oda,Ferreira:2014kpa,Andersen:2014oaa,Ferreira:2014exa,Braun:2014fua,Mitter:2014wpa,Grunfeld:2014qfa,Allen:2015qxa,Mueller:2015fka,Aoki:2015hsa}; see also the review paper~\cite{Andersen:2014xxa}.
We investigate the behavior of the RG flow of the four-Fermi coupling constant and we provide a detailed analysis of the phase diagram for the NJL model with an external magnetic field.

The remainder of this paper is organized as follows. The formulation is given in the next section.
Our results for structures of fixed points, the phase diagram, and the large-$N$ non-leading effects are presented in Section~\ref{section3}. 
We summarize and discuss our results in Section~\ref{section4}.
\section{ NJL model and its RG equations}\label{section2}
In this section, we briefly introduce the RG flow equation of the four-Fermi coupling constant and its structure in an external magnetic field at finite temperature and density.
The energy dispersion relation of a massless quark with electric charge $q$ in an external magnetic field ${\bvec B}=(0,0,B)$ is $E_n^2=(2n+1+s)|qB|+p_z^2$, where $s=\pm 1$ is the Zeeman splitting due to the interaction between the spin of a quark and $B$. 
The quantum number $n=0,1,2,\cdots$ is the Landau level and in the particular case where $n=0$ and $s=-1$, this is called the Lowest Landau level (LLL).
In the case of LLL, the dynamics of the quarks can be effectively described as a $1+1$ dimensional system.

We employ the following truncated effective action with ${\rm U}_{\rm L}(1)\times {\rm U}_{\rm R}(1)$ chiral symmetry in the Euclidean space,
\begin{align}
\nonumber
\Gamma_\Lambda=\int ^\beta _0d\tau \int d^3x
&\left[{\bar \psi}({\Slash \partial}+\mu+q{\Slash A})\psi \right. \\
\label{effectiveaction}
						&\left. -\frac{G_\Lambda}{2}\{ ({\bar \psi}\psi)^2 + ({\bar \psi}i\gamma_5\psi)^2\} \right],
\end{align}
where the external vector potential $A_\mu$ is defined to give ${\bvec B}=(0,0,B)={\rm rot}~{\bvec A}$.
The external field $A_\mu$ has several representations due to the gauge degrees of freedom, e.g., the symmetric gauge:~$A_\mu=(0,By/2,-Bx/2,0)$, or the Landau gauge:~$A_\mu=(0,0,Bx,0)$.

The effective action is governed by the Wetterich equation~\cite{Wetterich:1992yh,Morris:1993qb}, which in our case reads, 
\begin{align}
\partial _t \Gamma_\Lambda[\psi,{\bar \psi}]
=-{\rm Tr}\left[ \frac{\partial _tR_\Lambda}{\Gamma^{(1,1)}_\Lambda+R_\Lambda} \right],
\end{align}
where $\Gamma _\Lambda^{(i,j)}$ denotes the $i$-th ($j$-th) left (right)-hand side derivative of the effective action $\Gamma_\Lambda$ with respect to $\psi$ (${\bar \psi}$).
The cut-off profile function $R_\Lambda$ controls the shell momentum integration, thereby realizing the coarse-graining. 
We employ the $3d$ optimized cut-off function~\cite{Litim:2001up,Litim:2006ag},
\begin{align} 
R_\Lambda({\bvec p})
	=i{\Slash {\bvec p}}\left( \frac{\Lambda}{|{\bvec p}|} -1\right) \theta (1-|{\bvec p}|/\Lambda).
\end{align}
The momentum mode integral at finite temperature in the external magnetic field takes the following form,
\begin{align}
2\int\frac{d^4p}{(2\pi)^4} \to T\sum_{m=-\infty}^{\infty}\frac{|qB|}{2\pi}\sum_{l=0}^\infty \alpha_l \int \frac{dp_z}{2\pi},
\end{align}
where $m$ is the Matsubara mode number, and the factor 2 on the left-hand side and $\alpha_l=2-\delta_{l,0}$ on the right-hand side are the spin-degeneracy factors.
We also rewrite the Landau level as $2n+1+s\equiv 2l$ with $l=0,1,2,\cdots$.

Details of the derivation of the RG flow equations in this system were described previously~\cite{Fukushima:2012xw}.
The RG equations for the effective action (\ref{effectiveaction}) are reduced as follows:
\begin{align}\label{rgeqfinb}
\partial _t g&=-2g
		+g^2\left( 4{N_{\rm c}}J_0({\tilde T},{\tilde \mu},{\tilde B})-J_1({\tilde T},{\tilde \mu},{\tilde B})\right),\\
\partial _t {\tilde T}&={\tilde T},\\
\partial _t {\tilde \mu}&={\tilde \mu},\\
\partial _t {\tilde B}&=2{\tilde B},
\end{align}
where $g$ is the dimensionless rescaled four-Fermi coupling constant $G_\Lambda \Lambda^2/2\pi^2$, ${\tilde T}$, ${\tilde \mu}$ and ${\tilde B}$ are dimensionless external parameters, and $\partial _t$ denotes the derivative with respect to the dimensionless cut-off scale $t=\log (\Lambda_0/\Lambda)$.
The threshold functions $J_0$ and $J_1$ are defined by
\begin{align}
\nonumber
&J_0({\tilde T},{\tilde \mu},{\tilde B})= \frac{|q{\tilde B}|}{4}\sum _{l=0}^{\lfloor \frac{1}{2|q{\tilde B}|}\rfloor}\alpha_l \sqrt{1-{2l|q{\tilde B}|}}\\
&~~~~~~~~\times\left\{1-n_+ -n_- -\del_t( n_+ + n_-) \right\} ,\\
\nonumber
&J_1({\tilde T},{\tilde \mu},{\tilde B})= \frac{|q{\tilde B}|}{4}\sum _{l=0}^{\lfloor \frac{1}{2|q{\tilde B}|}\rfloor}\alpha_l \sqrt{1-{2l|q{\tilde B}|}}\\
\nonumber
&\times \left\{ \frac{1}{(1+{\tilde \mu})^2}\left( \frac{1}{2}-n_+\right)	+\frac{1}{(1-{\tilde \mu})^2}\left( \frac{1}{2}-n_-\right) \right. \\
&~~~~~~~~~~~~~~-\left. \frac{1}{1+{\tilde \mu}}\del_t n_+-\frac{1}{1-{\tilde \mu}}\del_t n_- \right\}.
\end{align}
The Gauss symbol $\lfloor x \rfloor$ denotes the greatest integer that is less than or equal to $x$.
The Fermi-Dirac distribution functions $n_\pm$ are defined as
\begin{align}\label{fermidis}
n_\pm=\frac{1}{e^{\beta (\Lambda \pm \mu)}+1}=\frac{1}{e^{{\tilde \beta} (1 \pm {\tilde \mu})}+1}.
\end{align}
The factor $4$ and $N_{\rm c}$ on the right-hand side of Eq.~(\ref{rgeqfinb}) denote the number of degrees of freedom for the spinor and the color of the fermionic fields, respectively.
The large-$N$ leading approximation neglects the quantum corrections corresponding to the term $J_1$ in Eq.~(\ref{rgeqfinb}).

\section{Results}\label{section3}
\subsection{Fixed Point Analysis}
In this subsection, we analyze the structure of the ultraviolet fixed point (UVFP) $g^\ast$, which satisfies $\beta(g^\ast)=0$ for the $\beta$ function of $g$ in Eq.~(\ref{rgeqfinb}). 
The four-Fermi coupling constant $g$ corresponds to the chiral susceptibility $\langle ({\bar \psi}\psi)^2 \rangle$.
Therefore, the RG flow equation of $g$ with the initial value $g_0>g^\ast$ diverges at a critical scale, where D$\chi$SB with the second order phase transition turns on. 

In the case of strong $B$ where the LLL approximation is effective with vanishing $T$ and $\mu$, the beta function of $g$ is given by
\al{
\beta_g=-2g +N_{\rm c}g^2 |q{\tilde B}|,
} 
which implies that the UVFP and the Gaussian fixed point ($g^{\ast}=0$) will become degenerate at large values of $t$.
This property is due to the dimensional reduction~\cite{Fukushima:2012xw}.
In fact, the RG flow of $g$ always diverges when $g_0>0$.
The thermal and density effects resolve the degeneracy of the fixed points. 
We investigate the dependence of the position of the UVPF on the temperature, chemical potential, and external magnetic field.
\begin{figure}
  \centerline{\hbox{
\includegraphics[width=80mm]{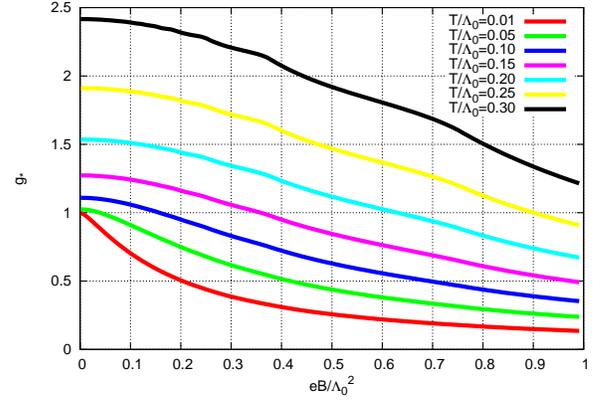}
    }}
\caption{Dependence of the position of the UVFP $g^\ast$ on the external magnetic field at finite temperature with the vanishing density in the large-$N$ leading approximation.}
\label{ebtfixedpoint}
\end{figure}
\begin{figure*}
  \centerline{\hbox{
\includegraphics[width=165mm]{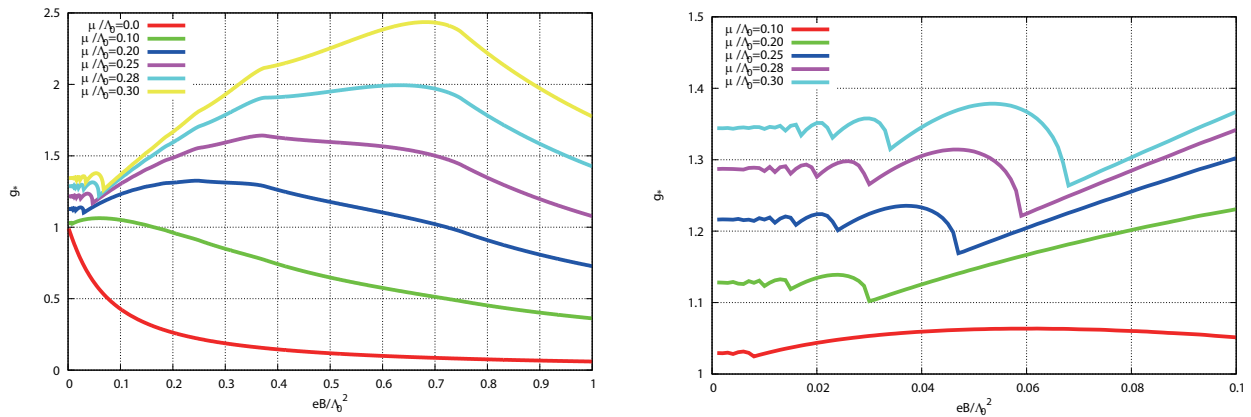}
    }}
\caption{Dependence of the position of the UVFP $g^\ast$ on the external magnetic field and finite density with fixed temperature ($T/\Lambda_0=0.0001$) in the large-$N$ leading approximation. The diagram on the right-hand side shows an enlargement of the small $B$ region.}
\label{ebtfixedpoint2}
\end{figure*}
\begin{figure}
  \centerline{\hbox{
\includegraphics[width=80mm]{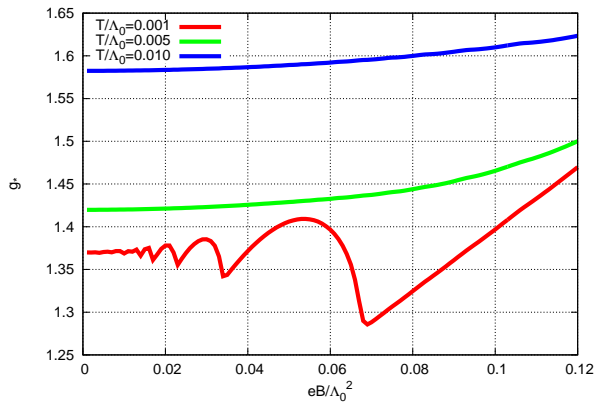}
    }}
\caption{Dependence of the position of the UVFP $g^\ast$ of Eq.~(\ref{largenrgeq}) on the external magnetic field with fixed chemical potential ($\mu/\Lambda_0=0.10$) and several temperatures in the large-$N$ leading approximation.}
\label{ebtfixedpoint3}
\end{figure}
The RG equation for the four-Fermi coupling constant in the large-$N$ leading approximation is given by
\begin{align}\label{largenrgeq}
\del_t g=-2g +4N_{\rm c}g^2J_0({\tilde T},{\tilde \mu},{\tilde B}),
\end{align}
where we set the following values: $N_{\rm c}=3$ and $q=(2/3)e$.
The UVFP $g^\ast$ of this equation is given by
\begin{align}
g^\ast =\frac{1}{6J_0({\tilde T},{\tilde \mu},{\tilde B})}.
\end{align}
The position of the UVFP with a finite temperature, finite external magnetic field, and vanishing chemical potential is shown in Fig.~\ref{ebtfixedpoint}.
We can see that $g^\ast$ decreases monotonically as $B$ increases for any temperature, and thus magnetic catalysis occurs.
This result agrees with previous studies of chiral effective models (e.g., ~\cite{Gusynin:1995nb}). 

\begin{figure}
  \centerline{\hbox{
\includegraphics[width=75mm]{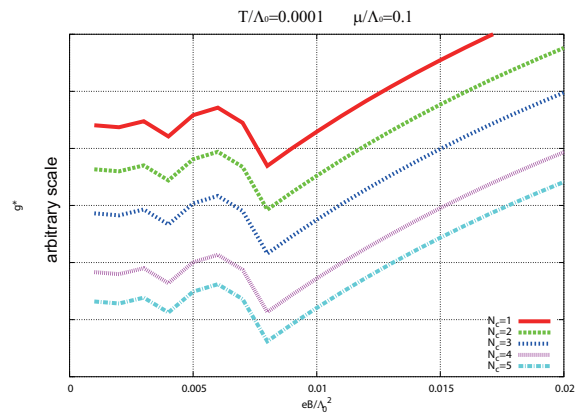}
    }}
\caption{Dependence of the position of the UVFP $g^\ast$ on various numbers for the color $N_{\rm c}$. The scale of the vertical axis is arbitrary.}
\label{ebtfixedpoint4}
\end{figure}

The change in the position of the UVFP caused by finite density and a finite external magnetic field with a fixed low temperature ($T/\Lambda_0=0.0001$) is shown in Fig.~\ref{ebtfixedpoint2}.
For the smaller $B$ region, the UVFP shown in the right-hand side panel of Fig.~\ref{ebtfixedpoint2} oscillates.
This behavior can be interpreted as the de Haas--van Alphen (dHvA) effect, which was observed in related studies, e.g., in color superconducting matter~\cite{Fukushima:2007fc}, holographic matter~\cite{Preis:2010cq,Preis:2012fh}, and in the NJL model with the mean field approximation~\cite{Inagaki:2003yi}.   
Thus, the oscillatory behavior is derived from the processes at each Landau level $2|qB|n$ that cross the Fermi surface $\mu^2$.
The end point of the oscillation corresponds to the case where the first Landau level ($n=1$) just overlaps with the Fermi surface.
Therefore, only the LLL can make a contribution beyond the oscillatory region.
For example, in case where $\mu/\Lambda_0=0.2$ the oscillation ceases at $eB/\Lambda_0^2\simeq 0.03$.
These values satisfy the relationship $2|qB|=2|\frac{2}{3}eB|\simeq \mu^2$.
The dHvA effect disappears at higher temperature (see Fig.~\ref{ebtfixedpoint3}).
The Landau level and the Fermi surface do not depend on the numbers of the flavor and color, and thus this phenomenon is not affected by them, as shown in Fig.~\ref{ebtfixedpoint4}.

\begin{figure*}
  \centerline{\hbox{
\includegraphics[width=160mm]{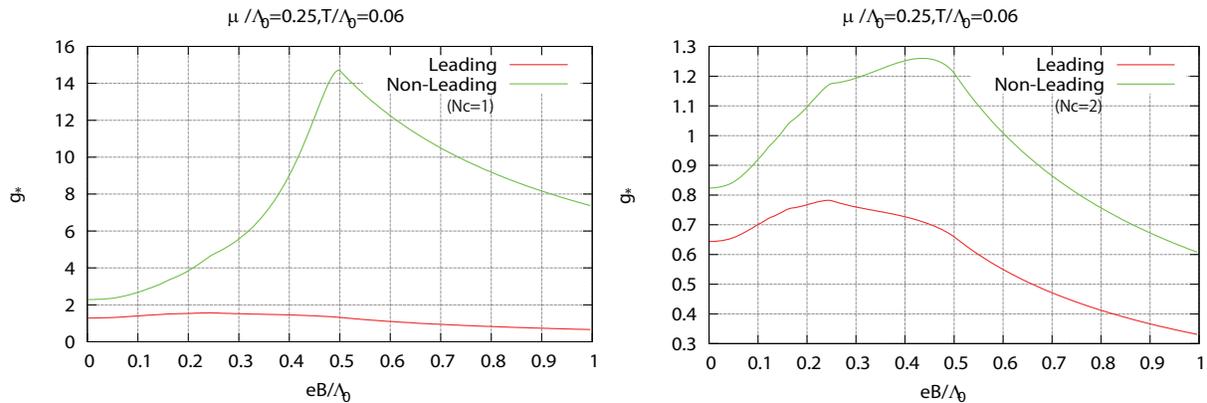}
    }}
\caption{Comparison of the position of the UVFP $g^\ast$ in the large-$N$ leading case and non-leading case with $N_{\rm c}=1$~(left) and $N_{\rm c}=2$~(right).}
\label{fixedpoints}
\end{figure*}
\begin{figure*}
  \centerline{\hbox{
\includegraphics[width=160mm]{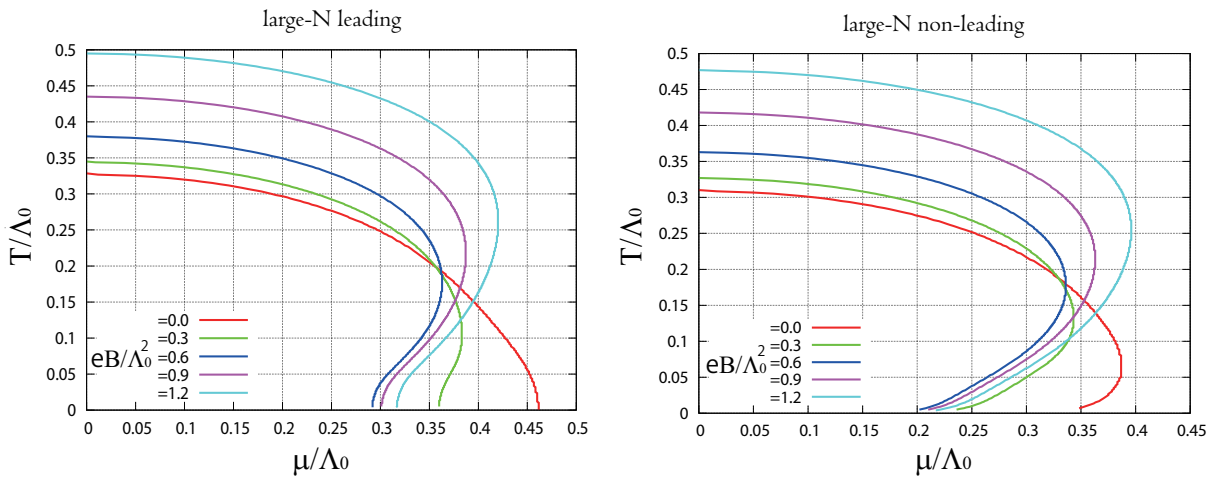}
    }}
\caption{Chiral phase diagram of the ${\tilde \mu}$-${\tilde T}$ plane. The figures on the left and right are the large-$N$ leading case and non-leading case, respectively.}
\label{phasediagram}
\end{figure*}

After the oscillation ceases, the UVFP $g^\ast$ increases with $B$, which means that inverse magnetic catalysis occurs.
The inverse magnetic catalysis at low temperature is reproduced qualitatively by analyzing the fixed point structure~\cite{Preis:2010cq,Preis:2012fh,Fukushima:2012xw}. This phenomenon cannot be observed in a lattice simulation at present because it is affected by the so-called sign problem.

Moreover, we note that the UVFP decreases for larger values of $B$.
It appears that the effect of finite density decouples with a larger external magnetic field because the magnetic field at the peak is proportional to the chemical potential.
At present, there is no clear explanation of this behavior.

We investigate the impact of the large-$N$ non-leading corrections using Eq.~(\ref{rgeqfinb}), where the term $J_1$ is included.
A comparison of the position of the UVFP $g^\ast$ in the large-$N$ leading case and the non-leading case with $N_{\rm c}=1$ and $N_{\rm c}=2$ is shown in Fig.~\ref{fixedpoints}.
The large-$N$ non-leading corrections make the UVFP $g^\ast$ larger in the whole region of $B$.
However, the qualitative magnetic behavior of the UVFP does not change greatly after the inclusion of the large-$N$ non-leading effects.
\subsection{Phase diagram}
The phase diagrams for the ${\tilde \mu}-{\tilde T}$ plane with fixed external magnetic field are shown in Fig.~\ref{phasediagram}. 
We evaluate the RG equation of the inverse four-Fermi coupling constant in order to obtain the four-Fermi coupling constant at the infrared limit~\cite{Aoki:2015hsa}.
Note that the phase boundaries in these phase diagrams do not necessarily indicate the true chiral phase transition point.
Instead, these boundaries represent the instability of the chiral symmetric vacuum, which is the phase transition point only for the second order transition.
For the first order phase transition, the true transition point moves to the symmetric side.
We set the dimensionless rescaled inverse four-Fermi coupling constant to $1/g=0.182$ at the initial scale $t=0$, for which the RG flow at vanishing temperature and density shifts to the broken phase at the infrared scale.

First, we discuss the large-$N$ leading case.
In the high temperature region, the phase boundaries move to the higher temperature side as $B$ increases, i.e., magnetic catalysis occurs.
By contrast, inverse magnetic catalysis occurs in the low temperature region. 
The phase boundaries then move toward the lower density side as $B$ increases.
Next, for larger values of $B$, they turn back toward the higher density side.
This behavior is also observed in the mean-field approximation~\cite{Inagaki:2003yi}. 
It appears that this behavior is due to competing effects between magnetic catalysis and inverse magnetic catalysis.

Next, we discuss the large-$N$ non-leading effects. 
At low temperature, the phase boundary turns toward the lower density side, even with the vanishing $B$.
This occurs due to the singularity at the Fermi surface (see~\cite{Aoki:2015hsa}). 
At vanishing temperature and finite density, the non-leading correction becomes singular at $\mu =\Lambda$, so the non-leading effect becomes larger than the leading one in the low temperature region. 
As shown by Eq.~\eqref{rgeqfinb}, the non-leading term has a negative sign in the RG equation of the four-Fermi coupling constant, i.e., the non-leading effect makes the phase more symmetric.
At finite $eB$, we still see that the phase boundaries move back and forth as the external magnetic field increases in the low temperature and high density region.
Thus, the non-leading effects do not change the behavior attributable to the external magnetic field.

\section{Summary and Discussion}\label{section4}
In this study, we investigated the dependence of the UVFP on the four-Fermi coupling constant in the NJL model at finite temperature and density under an external magnetic field by using the FRG.
The UVFP decreases monotonically as the magnetic field increases at finite temperature and vanishing chemical potential, and thus magnetic catalysis occurs.
At finite chemical potential and a fixed low temperature, the UVFP oscillates depending on the external magnetic field due to the dHvA effect.
This effect vanishes at higher temperatures.
The UVFP increase as the magnetic field increases after the oscillatory region, which means that inverse magnetic catalysis occurs.
However, an even larger external magnetic field changes the inverse magnetic catalysis into magnetic catalysis.

We also investigated the chiral phase diagram.
Magnetic catalysis is observed at high temperature and low density.
However, at low temperature and high density, inverse magnetic catalysis occurs with a large external magnetic field.
For a larger external magnetic field, the phase boundaries move back to the large density side, and thus magnetic catalysis occurs.
The large-$N$ non-leading effects do not change the qualitative behavior of our system dramatically, although the phase boundary moves toward the symmetric side at low temperature and high density in $eB=0$ due to the singularity at the Fermi surface~\cite{Aoki:2015hsa}. 

In order to investigate the cut-off scheme dependence, we analyzed the system with the $1d$ optimized cut-off function~\cite{Kamikado:2013pya},
\begin{align}\label{1dcut}
R_\Lambda(p_z)&=i p_z \gamma_z\left( \frac{\Lambda}{|p_z|}-1\right) \theta(1-|p_z|/\Lambda).
\end{align}
The threshold functions are shown in Appendix~\ref{threshold}.
We found that the qualitative behaviors do not change, such as the dependence of the UVFP on thermal effects and the shapes of the phase boundaries.
Clearly, the values of the critical temperature, density, and external magnetic field change because the NJL model is itself an unrenormalizable theory.
It appears that the behaviors of the UVFP and the phase diagram determined in this study are quite stable relative to the cut-off profile.

We also comment on the momentum-dependent coupling constant, i.e., the non-local vertex $G(p)$. The non-local vertex is partly included through the large-$N$ non-leading diagrams, and thus they can be considered to represent some of the fluctuations in mesons~\cite{Fukushima:2012xw}. The large-$N$ leading effect is much larger than the non-leading effects, so our results only change slightly. After including the momentum-dependence of the four-Fermi coupling constant, the phase boundary is expected to move toward lower temperature and density because the chiral symmetry tends to be restored by mesonic fluctuations.

We hope that these analyses motivate more elaborate studies in the future using the re-bosonization method~\cite{Aoki:1999dw,Gies:2001nw,Gies:2002hq,Floerchinger:2009uf,Braun:2014ata} or the weak solution method~\cite{Aoki:2014ola}.
Indeed, more precise analyses should be performed.

\subsection*{Acknowledgments}
We thank Motoi Tachibana and Daisuke Sato for fruitful discussions and comments, and Hokuriku--Shin-etsu Winter School, which was supported by the Yukawa Institute for Theoretical Physics (YITP-S-13-06).
M. Y. was supported by a Grant-in-Aid for JSPS Fellows (No.~25-5332).
K-I. A. was supported by a JSPS Grant-in-Aid for Challenging Exploratory Research (No.~25610103).
\begin{appendix}
\section{Threshold function with the $1d$ optimized cut-off function}\label{threshold}
We give the threshold functions using the $1d$ optimized cut-off function~(\ref{1dcut}) in the RG equation of the four-Fermi coupling constant. 
We have
\al{\nonumber
&I_0({\tilde T},{\tilde \mu},{\tilde B})
=\frac{|q{\tilde B}|}{4} \left[ -1 +\frac{\zeta(\frac{3}{2},\frac{1}{2|e{\tilde B}|})}{\sqrt{2}|e{\tilde B}|^{3/2}} \right.\\
&~~~\left. - \sum_{l=0}^\infty \alpha_l \left\{ \frac{\left( n_+ +n_- \right)}{{\tilde \epsilon}_l^3} -\frac{\del_t(n_+ +n_-)}{{\tilde \epsilon}_l} \right\} \right], \\
\nonumber
&I_1({\tilde T},{\tilde \mu},{\tilde B})
=\frac{|q{\tilde B}|}{4}\sum_{l=0}^\infty \alpha_l
\left[ \frac{1}{{\tilde \epsilon}_l({\tilde \epsilon}^+_l)^2}\left(\frac{1}{2}-n_+ \right) \right.\\ 
&~~~\left.+\frac{1}{{\tilde \epsilon}_l({\tilde \epsilon}^-_l)^2}\left( \frac{1}{2} - n_-\right) 
-\frac{\del_t n_+}{{\tilde \epsilon}_l^+}  - \frac{\del_t n_-}{{\tilde \epsilon}_l^-} \right],
}
where $\zeta(s,x)$ is the Hurwitz zeta function, $n_\pm$ are given in Eq.~(\ref{fermidis}), and ${\tilde \epsilon}_l^\pm ={\tilde \epsilon} _l \pm {\tilde \mu}$ with ${\tilde \epsilon}_l^2=1 +2l|q{\tilde B}|$.
The RG equation of the four-Fermi coupling constant is obtained by replacing $J_0$ and $J_1$ with $I_0$ and $I_1$, respectively.

\end{appendix}

\bibliography{refs}

\end{document}